\documentclass[]{spie}  

\usepackage{amsmath,amsfonts,amssymb}
\usepackage{graphicx}
\usepackage[colorlinks=true, allcolors=blue]{hyperref}
 \usepackage[separate-uncertainty=true,range-phrase=--]{siunitx}
\usepackage{geometry}
\title{Spectrograph design for the Asgard/BIFROST spectro-interferometric instrument for the VLTI}

\author[a]{Sorabh Chhabra}
\author[b]{Michele Frangiamore}
\author[a]{Stefan Kraus}
\author[b]{Andrea Bianco}
\author[c]{Francisco Garzon}
\author[d]{John Monnier}
\author[a]{Daniel Mortimer}

\affil[a]{School of Physics and Astronomy, University of Exeter, Exeter, EX4 4QL, UK}
\affil[b]{INAF-Osservatorio Astronomico di Brera, Via Bianchi 46, I-23807 Merate (LC), Italy}
\affil[c]{Instituto de Astrof´ısica de Canarias, La Laguna, E-38200 Tenerife, Spain}
\affil[d]{Department of Astronomy, University of Michigan, 1085 S University Ave, West hall, Ann Arbor, MI 48109, USA}

\authorinfo{--}

\begin{document} 
\maketitle

\begin{abstract}
The BIFROST instrument will be the first VLTI instrument optimised for high spectral resolution up to R=25,000 and operate between 1.05 and 1.7 µm. A key component of the instrument will be the spectrograph, where we require a high throughput over a broad bandwidth. In this contribution, we discuss the four planned spectral modes (R=50, R=1000, R=5000, and R=25,000), the key spectral windows that we need to cover, and the technology choices that we have considered. We present our plan to use Volume Phase Holographic Gratings (VPHGs) to achieve a high efficiency $>$ 85\%. We present our preliminary optical design and our strategies for wavelength calibration. 
\end{abstract}


\section{INTRODUCTION}
\label{sec:intro}  

Long baseline optical interferometers bring revolution in high-angular resolution astronomy. With the help of VLTI infrastructure, which comprises 4 Unit Telescopes (UTs) and 4 Auxiliary Telescopes (ATs) of 8.4 m and 1.8 m respectively, interferometry enables a resolution down to sub-milliarcseconds.

BIFROST (\textbf{B}eam-combination \textbf{I}nstrument for studying the \textbf{F}ormation \& fundamental p\textbf{R}operties \textbf{O}f \textbf{S}tars and plane\textbf{T}ary systems) is an upcoming Very Large Telescope Interferometry (VLTI) visitor instrument as part of VLTI/Asgard instrument Suite\cite{bifrost_kraus,asgard_martinod,bifrost_mortimer,hi5_dandumontA,hi5_dandumontB,hi5_defrere,hi5_garreau,hi5_laugier,hi5_sanny,2018SPIE10701E..11I}. BIFROST will operate in either a Y+J or a H band mode. The major components that we need to design for BIFROST include the spectrograph. This paper provides details on the BIFROST spectrograph, which is motivated by the state-of-the-art instrument MIRC-X spectrograph design but with an additional provision for a high resolution (HR) arm.  MIRC-X\cite{anugu2020mirc} (Michigan InfraRed Combiner-eXeter) is a beam-combiner that combines the light from the six 1\,m telescopes at the CHARA Array and that operates in the JH band.

Section~\ref{sec:spectrograph_design} covers the conceptual design of the spectrograph. Section~\ref{sec:spectral_modes} outlines the main modes of the spectrograph. Section~\ref{sec:dispersing_elements} covers potential dispersing elements for both arms of the spectrograph. Section~\ref{sec:spectral_calibration} discusses our strategy for wavelength calibration. And last, section~\ref{sec:conclusion} concludes this paper with final remarks.

\section{Spectrograph design}
\label{sec:spectrograph_design}
\begin{figure}[h!]
    \centerline{\includegraphics[width=1\textwidth]{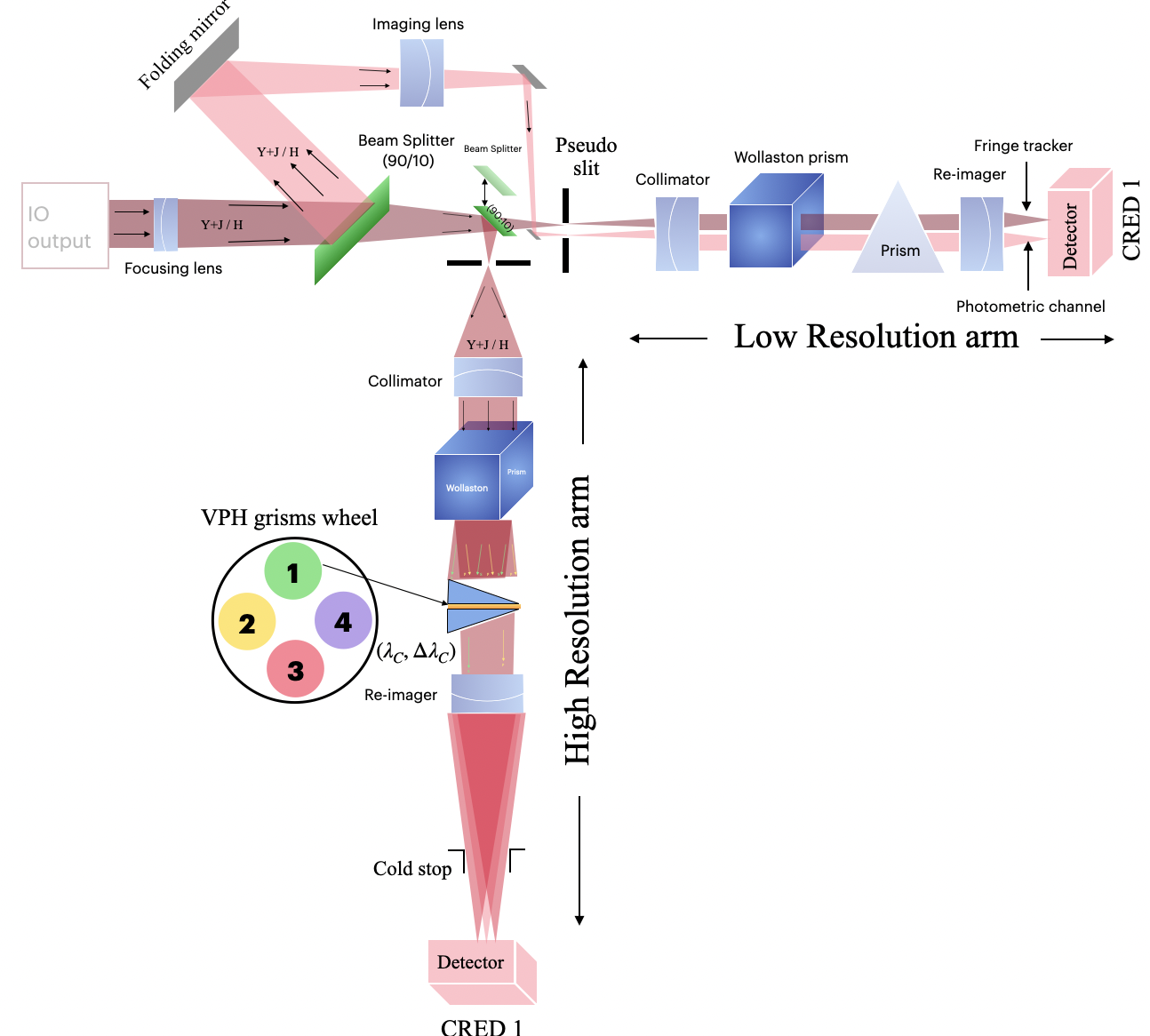}}
    \caption{\label{fig:spectrograph} Conceptual design of spectrograph for BIFROST comprises of two arms LR and HR (here for an All-In-One beam combiner, instead of an Integrated Optics device). }
\end{figure}

The conceptual design for the BIFROST spectrograph is shown in fig.~\ref{fig:spectrograph}, which is inspired by the MIRC-X\cite{anugu2020mirc} spectrograph. An important addition is the HR arm, which comprises various high-resolution spectral modes up to R~=~25,000. The output beam from the Integrated Optics (IO) chip will first pass through focusing optics and will form an image at a pseudo slit.  A beam splitter will split the light between the low and high resolution arms in a 10:90 ratio (with 10\% going to the low resolution arm). This can be removed to send 100\% of the starlight to the low resolution arm for observations of faint targets. Part of the beam gets split by a beam splitter (1/10) for its use as a photometric channel (just like in MIRC-X for fiber injection module). Each arm has optics for collimation and for reimaging, other components including Wollaston prism, dispersing element (prism for low resolution and VPH grism wheel for high resolution arm, covered in sec.\ref{sec:HR}), and C-RED One detector\cite{anugu2018mirc}. Wollaston helps get the best contrast in the fringes thus by providing feedback to Lithium Niobate LiNbO3 plate incidence angle of polarization\cite{bifrost_mortimer,anugu2020mirc}. For the initial implementation, we plan to use 320x256 pixel C-RED One cameras on both arms. C-RED One\cite{lanthermann2018astronomical,gach2016c} is an ultra-low noise Short Wavelength IR (SWIR) camera that uses a 320x256 pixels HgCdTe e-APD array with sub-electron noise and cut-off at 3.5$\mu$m.

\section{Spectral modes}
\label{sec:spectral_modes}
\begin{figure}[h!]
    \centerline{\includegraphics[width=0.9\textwidth]{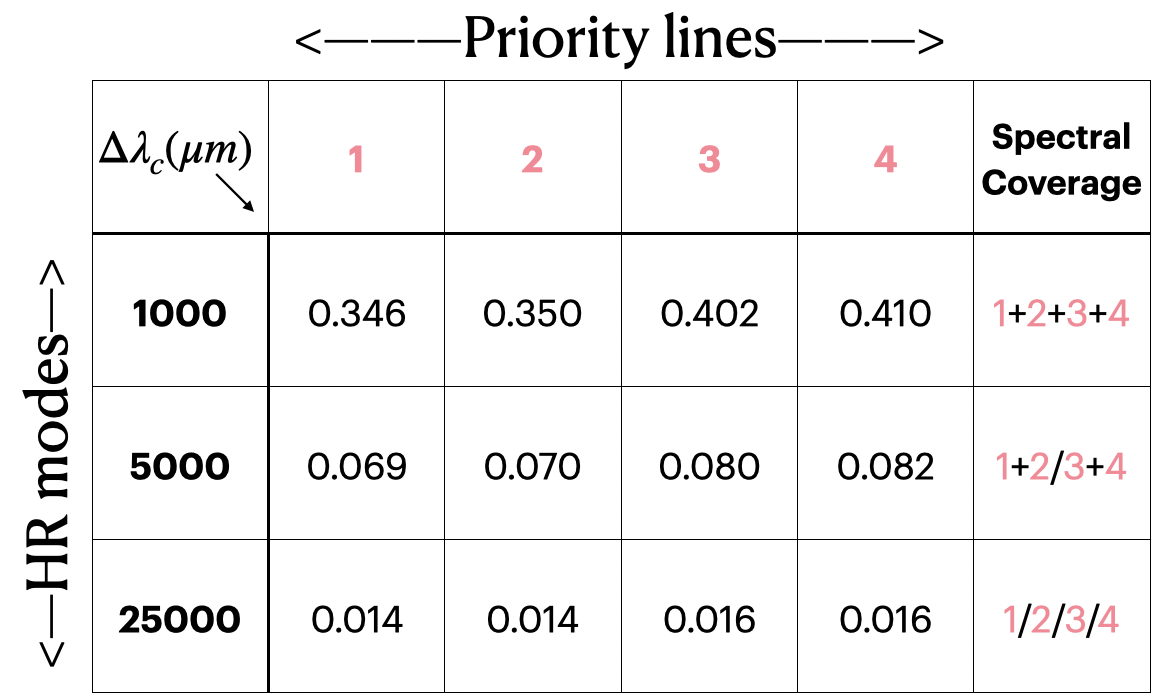}}
    \caption{\label{fig:spectral_modes} Spectral modes (bandwidth $\Delta\lambda_c$) and spectral coverage of medium, high and very high resolution modes for priority lines 1 to 4. Spectral bandwidth calculations are based on a C-RED One detector over a 320-pixel line}
\end{figure}

The BIFROST spectrograph design comprises two arms - a low resolution (LR) and high resolution (HR) arm (as shown in fig.~\ref{fig:spectrograph})\footnote{In Kraus et al.\cite{bifrost_kraus}, the LR~arm is referred to as YJH~arm and the HR~arm as YJ~arm, as we might chose a close filter with a lower cut-off wavelength on the arm with the R=1000, R=5000, and R=25,000 grating for Y and J-band to reduce the thermal background and to enable very long effective integration times.}. The LR arm comprises as dispersing elements a prism, with R50 to 100 (more details covered in section.~\ref{sec:LR}). The LR arm is meant to cover a large set of Brackett lines in the H band (shown in fig.~\ref{fig:spectral_band}). The HR arm comprises of three important spectral modes R~=~1,000, R~=~5,000, and R~=~25,000 for science lines in the Y and J-band, including HeI (priority line 1: $\lambda_c$ = \SI{1.083}{\micro\meter}), H-Pa$\gamma$ (priority line 2: $\lambda_c$ =\SI{1.094} {\micro\meter}), [Fe\,II] (priority line 3: $\lambda_c$ = \SI{1.257}{\micro\meter}) and H-Pa$\beta$ ( priority line 4: $\lambda_c$ = \SI{1.282}{\micro\meter}). All these lines cover a maximum of 300\,nm bandwidth in the Y+J band. For R~=~5,000 and R~=~25,000, it is impossible to cover the entire bandwidth within the C-Red ONE detector size of 320x256 (where a 320-pixel side is used as the spectral line and a 256-pixel side is used for the fringe encoding). To cover the most within the proposed design (from fig.~\ref{fig:spectral_modes}), a table lists probable spectral modes for the HR arm on BIFROST. For R~=~1,000, it is clear that the entire 300\,nm bandwidth in the Y+J bands covers all four priority lines within a single setting onto our detector. However, for R~=~5,000, either Y, i.e.\ target 1 \& 2 or J, i.e.\ target 3 \& 4, can be acquired simultaneously within the detector itself. At last, R~=~25,000 can only cover one target at a time with a maximum bandwidth of $\Delta\lambda_c$ =16\,nm.

\section{Dispersing elements}
\label{sec:dispersing_elements}
\begin{figure}[h!]
    \centerline{\includegraphics[width=0.9\textwidth]{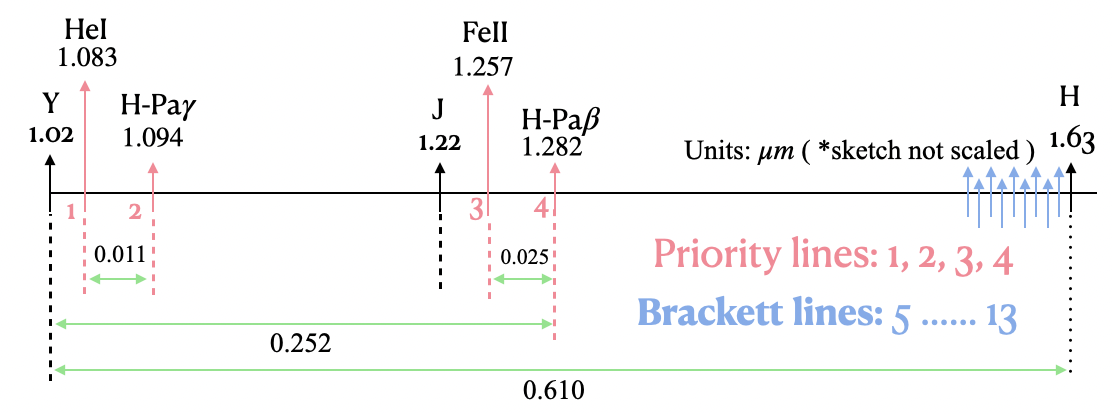}}
    \caption{\label{fig:spectral_band} Spectral band for BIFROST spectrograph covering important science targets lines 1 to 4}
\end{figure}

The dispersing element is the most central element of a spectrograph. Critical considerations for the choice of the dispersing elements are parameters like - refraction or diffraction, resolution ($R = \lambda/\Delta\lambda$), and efficiency ($\eta$). Typical choices for the dispersing elements include prisms, diffraction gratings, grisms, echelle gratings, binary gratings etc. Each one has its advantages and disadvantages. Prism can be the most suitable fit for applications that require only zero-order spectra but are not suitable for applications that demand high spectral resolution within small space of few cm. For the BIFROST spectrograph, we are planning to work with a prism for the LR arm (sec.~\ref{sec:LR}), and Volume Phase Holographic (VPH) gratings and binary gratings for the HR arm (sec.~\ref{sec:HR}). Fig.~\ref{fig:spectral_band} illustrates the working spectral range for the spectrograph, including priority lines 1 to 4. 

\subsection{LR arm}
\label{sec:LR}

As the name suggests, the LR arm only handles low spectral modes of BIFROST. A prism will be used as dispersing element for this arm, whose resolution will be around R~=~50-100, where the precise value still needs to be determined based on considerations on the interferometric field of view. Higher resolution can provide a larger interferometric field of view and better estimates of group delays, but at the cost of signal-to-noise ratio / sensitivity.  The LR arm performs two parameter evaluations for BIFROST - fringe tracking and (and in the case of an all-in-one beam combination design) also recording of the photometric signal (not required in case of IO chips). Since BIFROST operates either in Y+J or H band, the LR arm needs to be optimised for both input bands.

\subsection{HR arm}
\label{sec:HR}

As the name implies, the HR arm handles high spectral modes of BIFROST (R~=~1,000, R~=~5,000 and R~=~25,000). 

\begin{figure}[h!]
    \centerline{\includegraphics[width=0.6\textwidth]{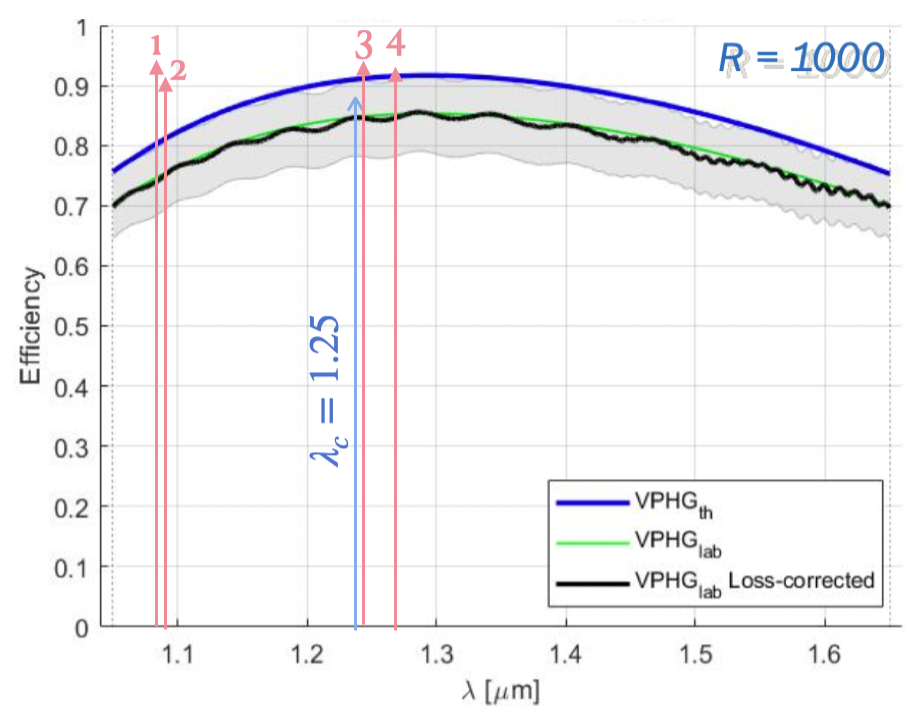}}
    \caption{\label{fig:R1000} RCWA (Rigorous coupled wave analysis) output for R~=~1,000 spectral mode at $\lambda_c = 1.25\mu$m, covering theoretical, laboratory and loss-corrected lab efficiency curve.}
\end{figure}

\begin{figure}[h!]
    \centerline{\includegraphics[width=0.6\textwidth]{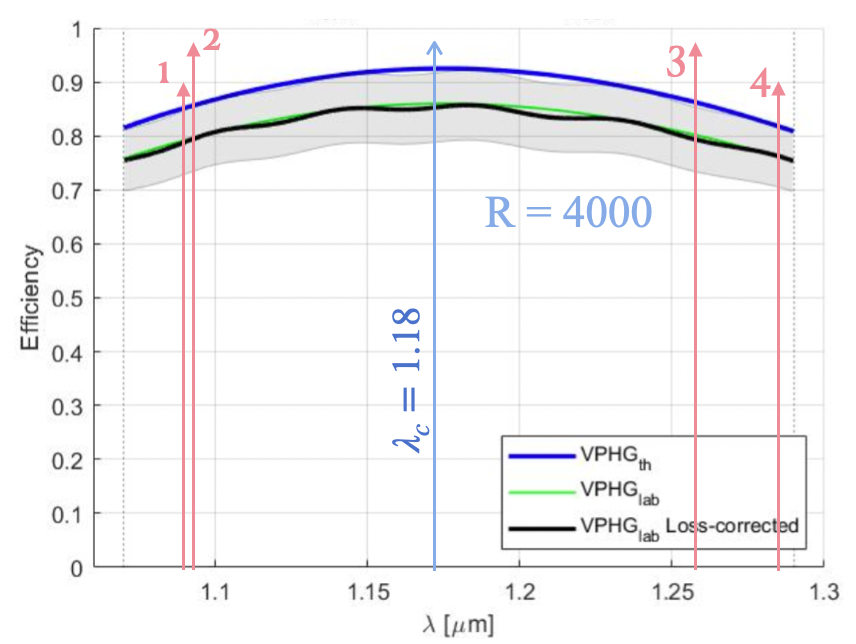}}
    \caption{\label{fig:R4000} RCWA output for R~=~4,000 spectral mode at $\lambda_c = 1.18\mu$m, covering theoretical, laboratory and loss-corrected lab efficiency curve.}
\end{figure}

For R~=~1,000 and R~=~5,000, VPH transmission grisms have been selected (analysis covered in next section.~\ref{sec:VHR_analysis}) and analysed by the team at Instituto Nazionale di Astrofisica (INAF), Italy. Transmission grating helps limit the complexity of the entire optics rather than reflective optics, which generally demand more optical elements. Special proprietary software has been in use to evaluate VPH grism efficiency. A closed analytical form of the Kogelnik equation is found to be sufficient for throughput calculation  and has been worked out to approximate diffraction efficiency. From our initial run for R~=~1,000 with a fixed focal length of 100\,mm, fig.~\ref{fig:R1000} shows the efficiency curve w.r.t. theoretical, lab corrected, and loss-corrected. It seems promising, with mean efficiency close to $\> 79.6\%$ and $\> 70\%$  for wavelength up to 1.6\,$\mu$. Parameters for the R~=~1000 VPH grating include thickness ($d$) of 60\,$\mu$, refractive index modulation ($\Delta n$) within VPH of 0.0118, incident angle of $3.68^{\circ}$, line density (l/mm) of 154. Similarly, for the R~=~5,000 mode, our initial test with a fixed focal length of 100 mm gives the best-optimised results for R~=~4,000 (fig.~\ref{fig:R4000} with 615 lines per mm and 21.29 incident angle. Thus we are planning to make a similar attempt again in future for R~=~5,000 resolution, also while considering other focal lengths.

We refer to R~=~25,000 as very high-resolution mode. We aim to optimise the VPH grating with high diffraction efficiency over a small bandwidth range. We will employ different grisms for different central bandwidth values to cover most of our priority lines. Using simple echelle gratings seems a better solution because it can bring high efficiency and cover the entire spectral range with just one grating\cite{wildi2017nirps}. However, reflective grating (as previously stated) for very high-resolution mode and transmission gratings for other resolution modes brings multi-fold complications like multiple reflection mirrors, detector space constraints, etc. Also, our detector is only $320\times 256$ pixels with $24\mu$m pixel pitch, thus making it difficult to use a VPH grism wheel along with the reflective dispersing element. Therefore, the following section \ref{sec:VHR_analysis} covers our efforts on VPH grisms for R~=~25,000 and the subsequent section \ref{sec:proposed} talks more about initial exploration of alternatives for high-efficiency grating for R~=~25,000.

\subsection{VPH grism characterisation and  performance analysis}
\label{sec:VHR_analysis}
\begin{figure}[h!]
    \centerline{\includegraphics[width=0.6\textwidth]{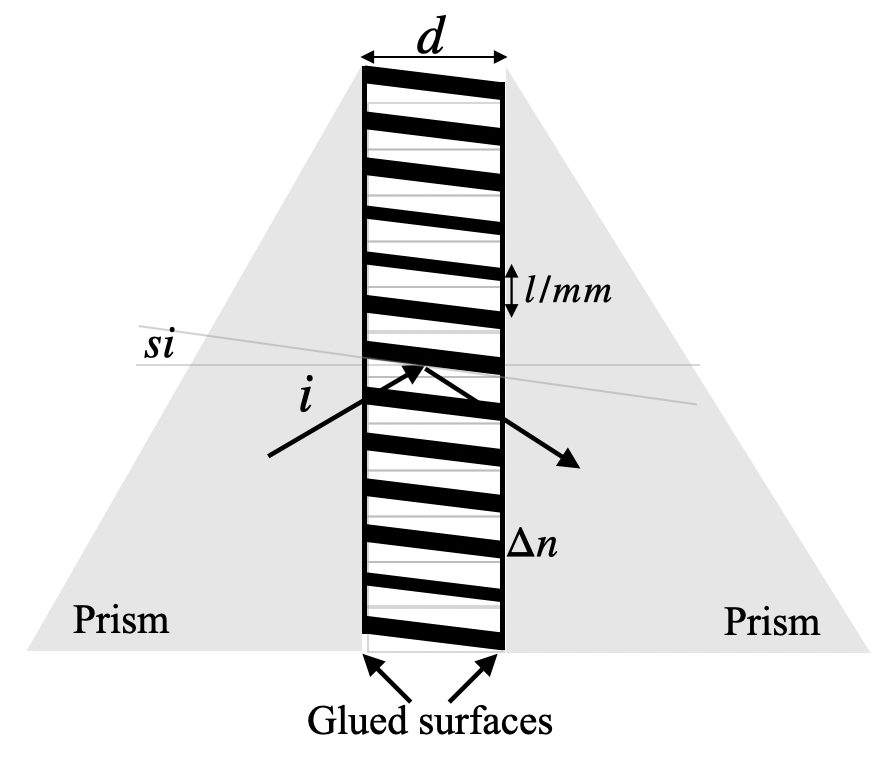}}
    \caption{\label{fig:vph_grism} 
     A sketch of VPH grism with the characterising parameters.}
\end{figure}

Five parameters can fully characterise VPH gratings: fringe orientation (si), fringe density (l/mm), Bragg angle (or incident angle $i$), grating depth ($d$), and refractive index modulation ($\Delta n$). All these parameters need to be considered for fabrication, wavelength selection and to optimise the corresponding efficiency in the desired order. Fig.~\ref{fig:vph_grism} shows the sketch diagram of VPH grisms along with the characterising parameters.  

\begin{figure}[h!]
    \centerline{\includegraphics[width=0.6\textwidth]{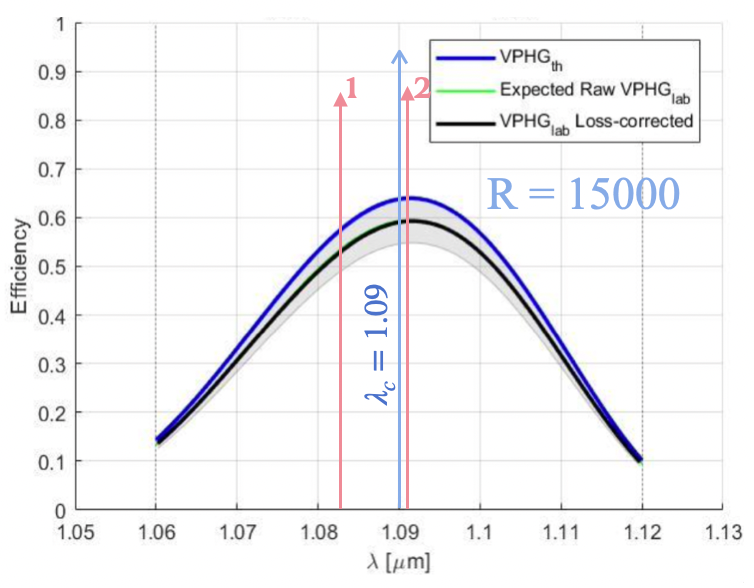}}
    \caption{\label{fig:R15000} RCWA output for R~=~15,000 spectral mode at $\lambda_c = 1.09\mu$m, covering theoretical, laboratory and loss-corrected lab efficiency curve.}
\end{figure}

Our test run on VPH grism at $f = 100 mm$ for R~=~15,000 at $\lambda_c = 1.09$ (as shown in fig.\ref{fig:R15000}) turns out to be less efficient with mean efficiency $<55\%$ with $\Delta\lambda_c = 0.02\mu$. Given that we need to cover only a small bandwidth, these gratings are still a viable solution, but we consider in parallel also alternative technologies, as will be discussed in the next section.

\subsubsection{Proposed alternative solution}
\label{sec:proposed}
\begin{figure}[h!]
    \centerline{\includegraphics[width=0.5\textwidth]{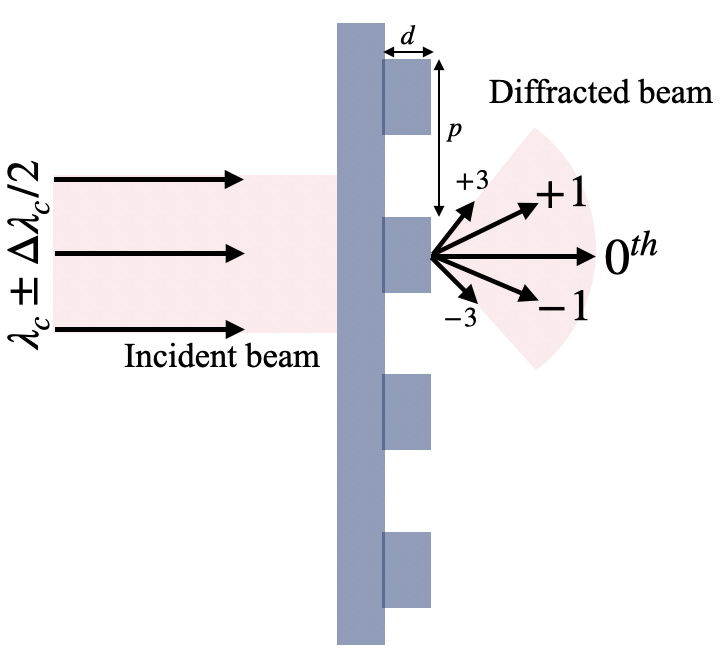}}
    \caption{\label{fig:binary}Diffracted light through binary transmission grating with period $p$ and depth $d$ for incident light of central wavelength $\lambda_c$ and bandwidth $\Delta\lambda_c$.}
\end{figure}

As an alternative technology that could provide very high spectral dispersion with high efficiency, we consider a transmission binary grating. Binary gratings proved to be very efficient for space missions in terms of high throughput, including Sentinel-4 (NIR 750-775, $\eta>70\%$) and Sentinel-5 (NIR 755-773, $\eta>85\%$).  More details about successful missions can be found in a review paper\cite{guldimann2015overview}. Binary surface relief gratings design approximate for binary gratings (as shown in fig.~\ref{fig:binary}) enable efficient coupling into first-order diffraction. Binary gratings can be fabricated to sub-wavelength order, giving high throughput even for small wavelengths. The first order high diffraction efficiency is directly linked with period ($p$) to wavelength ratio and period to depth ratio\cite{zeitner2012high}. Reflective binary gratings exist, which prove to be very efficient up to $1.09\mu$m with efficiency reaching up to $100\%$ for bandwidth up to $\sim 10$\,nm. Also, for larger bandwidth up to 120\,nm, efficiency is always  $>92\%$\cite{zeitner2012high}. Thus, we will investigate further transmission binary gratings for spectral ranges Y, J and H. Another critical parameter is polarisation independence which cannot be neglected in the case of binary gratings. Binary gratings are sensitive towards incoming light polarisation, and efforts are already in place to make it polarisation independent\cite{zeitner2012high,zhu2020design}.

\section{Spectral calibration}
\label{sec:spectral_calibration}
\begin{figure}[h!]
    \centerline{\includegraphics[width=0.7\textwidth]{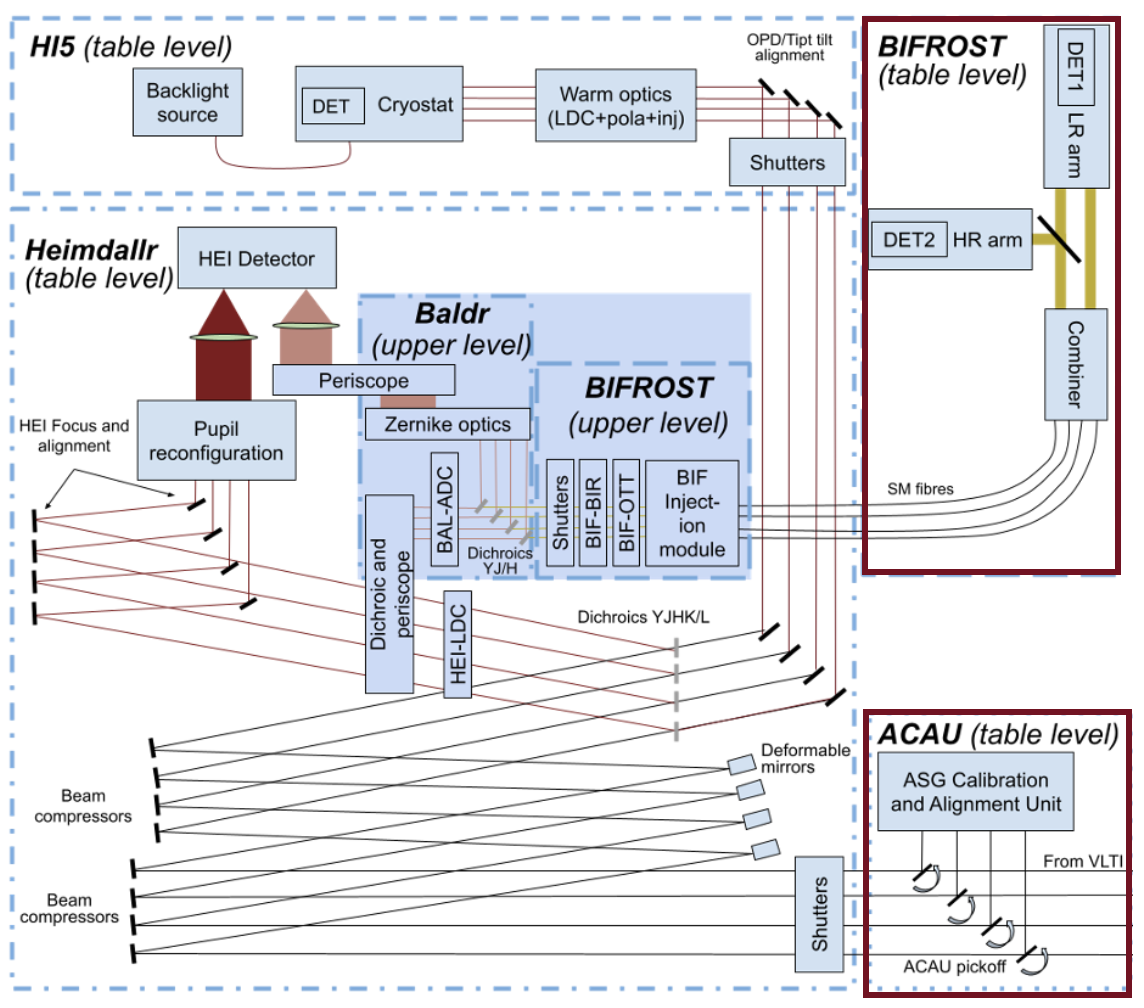}}
    \caption{\label{fig:asgard_cal} ASGARD\cite{asgard_martinod} calibration source location (bottom right) with reference to BIFROST location (top right) }
\end{figure}

Spectral calibration plays a critical role in a solid understanding of the apparatus or the elements dispersing behaviour before the light reaches the detector. It helps provide accurate dispersion figures for each central wavelength under test. Thus, it ultimately helps achieve the required accuracy before every night of observation. The BIFROST design demands an accuracy up to $1$\,km s$^{-1}$ (Kraus et al.\cite{bifrost_kraus}), so that the central wavelength of each spectral channel can be defined with sub-channel accuracy. Fig.~\ref{fig:asgard_cal} shows the location of the Asgard/BIFROST\cite{asgard_martinod} instrument calibration source location (bottom right), w.r.t. BIFROST instrument location (top right). 

Priority lines 1 \& 2 and similarly Brackett lines in the H band (from fig.~\ref{fig:spectral_band}) reside very close to each other. These delicate features demand three critical requirements from the calibration source side (i) Sharp well-characterised lines within the range of 1000-1700 nm, (ii) Source stability and (iii) Good lifetime. Fabry–Pérot interferometers or Etalons are well-studied and established solutions for wavelength calibration. For instance, MIRC-X\cite{anugu2020mirc} has used this setup and achieves routinely 0.25\% relative accuracy. However, for BIFROST an absolute wavelength calibration would be desirable. CRIRES\cite{aldenius2008wavelength} at ESO uses a Fourier Transform Spectrometer (FTS) following Th-Ar hollow cathode lamp (HCL), which provides much more stability and is highly reliable for absolute calibration. X-shooter\cite{kerber2008wavelength} later did a rigorous analysis using various combinations of gases like Ne, Ar, Kr and Xe and derived spectral lines table. Thus, various combinations of gases can be used per requirements. We will decide on a suitable solution following further analysis.

\section{Conclusion}
\label{sec:conclusion}
This paper covers the conceptual design for the BIFROST spectrograph with its high-resolution spectral modes R~=~1,000, R~=~5,000, and R~=~25,000 for four high-resolution priority lines HeI, H-Pa$\gamma$, [Fe\,II] and H-Pa$\beta$. The dispersing element for the LR arm will be a prism (between R$\approx$50-100), while the HR arm will use VPH gratings to achieve R~=~1,000 and R~=~5,000. A VPHG might also be a suitable solution for the very high resolution (R~=~25,000) mode, although we consider also binary gratings as alternative technology. For spectral calibration, a source like a Th-Ar HCL tube can be employed for absolute calibration.

\acknowledgments 
We acknowledge support from ERC Consolidator Grant (Grant Agreement ID 101003096) and STFC Consolidated Grant (ST/V000721/1). We thank our partner institute, INAF, Italy, for performing RCWA calculations.

\bibliography{report} 
\bibliographystyle{spiebib} 

\end{document}